\documentclass[12pt,preprint]{aastex}
\usepackage{graphicx}
\usepackage{longtable}
\usepackage{latexsym}
\usepackage{amssymb}
\usepackage{lscape}

\begin{document}

\title{Capturing a star formation burst in galaxies infalling onto the cluster Abell 1367 }
\author{G. Gavazzi$^1$, L. Cortese$^1$, A. Boselli $^2$, 
J. Iglesias-Paramo$^2$, J. M.  V\' \i lchez$^3$ \& L. Carrasco$^{4,5}$}

\affil{
$^1$ Universit\`a degli Studi di Milano - Bicocca, P.zza delle scienze 3,
20126 Milano, Italy\\ 
$^2$ Laboratoire d'Astrophysique de Marseille, BP8, Traverse du Siphon, F-13376 Marseille, France\\
$^3$ Instituto de Astrof\' \i sica de Andaluc\' \i a, CSIC, Apdo. 3004, 18080,
Granada,  Spain\\
$^4$Instituto Nacional de Astrof\' \i sica, Optica y Electr\'onica,
Apartado Postal 51. C.P. 72000 Puebla, Pue., M\'exico\\
$^5$Observatorio Astron\'omico Nacional, UNAM, Apartado Postal 877, C.P. 22860, Ensenada B.C., M\'exico}

\begin{abstract}
The discovery  of a striking astrophysical laboratory in the cluster of galaxies Abell 1367 
(Sakai et al. 2002) is confirmed 
with independent imaging and spectroscopic observations and further investigated in the present analysis.
Two giant and ten dwarf/HII galaxies, members to a group, are simultaneously undergoing a 
burst of star formation. Redshift measurements 
suggest that the group galaxies are in the process of falling into the cluster at very high speed.
We explore two possible mechanisms that could have triggered the 
short-lived stellar burst we are witnessing: 
the first, internal to the group itself, via tidal interactions among its members, 
the hypothesis favoured by Sakai et al. (2002); the second associated with the high velocity infall 
of the group galaxies into the cluster intergalactic medium. 
We present evidences in favour and against the two hypotheses. 

\end{abstract}


\section {Introduction}
As part of an $\rm H\alpha$ survey of nearby clusters of galaxies (Iglesias-Paramo et al. 2002)
we observed the field centered on NGC3860, projected near the X-ray center of the cluster Abell 1367 
(Donnely et al. 1998). 
Observations taken in the H$\alpha$ line provide the most direct probe of the current ($t$ $<$ 20 Myr) 
formation of 
massive ($m$ $>$ 10 M$_{\odot}$) stars (Kennicutt 1989). These observations provided us with a dozen 
star-forming galaxies around NGC3860, an exceedingly high density of such objects.\\
Meanwhile Sakai et al. (2001, 2002) independently observed the region with a similar observational setup, 
pointing out the uniqueness of the star forming properties in this particular area of the sky.\\
Redshift measurements obtained by the two groups independently
revealed that the star forming galaxies belong to a compact group with a recessional velocity
exceeding that of the cluster by $\rm \sim 1800 ~km~s^{-1}$. 
Sakai et al. (2002) speculate that the enhanced star formation was triggered 
internally to the group itself, by tidal interaction among its galaxies, 
independently from the presence of A1367.\\
Here we report\footnote{Based on observations taken at the INT, SPM, Loiano, and Calar Alto observatories.
The INT is part of the of the Isaac Newton group of telescopes (ING) located at the Roque de Los Muchachos
Observatory, La Palma, Spain. The SPM observatory
belongs to O.A.N de Mexico. The Loiano observatory belongs to the University of Bologna. The Calar Alto
observatory is operated by the Centro Astronomico Hispano Aleman.} on our $\rm H\alpha$ observations and on 
spectroscopy of the star forming galaxies.\\ 
The paper is arranged as follows. The observations and the data reduction procedures are described in Section 2. 
In Section 3 we try to characterize the nature of the group by sketching a possible 3-D picture. 
In Section 4 we discuss the interpretation proposed by Sakai et al. (2002) and we sketch
an alternative scenario for this unusual star formation event based on the interaction between 
the group and the cluster.
A Hubble constant of $H_o$ = 75 km s$^{-1}$ Mpc$^{-1}$ is used throughout this paper.

\section{Observations} 

\subsection{Imaging} 
As part of the $\rm H\alpha$ survey of A1367 (Iglesias-Paramo et al. 2002) we 
obtained one hour integration of the cluster central region using the Wide Field Camera (WFC) attached to the 
2.5 m Isaac Newton Telescope (INT) through a narrow band filter centered at $\sim$ 6725~\AA,
covering the redshifted $\rm H\alpha$ and [NII] lines.
The underlying continuum was taken through the (Sloan-Gunn) $r'$ broad band red filter with an exposure of 15 min.
Since the region of interest around NGC3860 was partly hidden by the gap between two chips of the camera, 
in march 2002 we obtained  
one additional 2 hours $\rm H\alpha$ exposure (20 min in the continuum) 
with the 2.1m telescope at San Pedro Martir (Mexico), with
a similar experimental setup. All images were obtained in photometric conditions
with a seeing of $\sim$ 1.8~arcsec.
The photometric calibration was obtained exposing the star Feige 34. 
The individual images were bias subtracted and flat-fielded using combinations of exposures of empty fields taken at 
twiglight and re-binned to a common pixel size.
After background subtraction the intensity in the combined OFF-band frames 
was normalized to that of the combined ON-band one using the flux ratio of several field stars.
The combined NET frame was obtained by subtraction of the OFF from the ON frame. 
The resulting stacked frames, corresponding to 3 hours (ON-band) and 35 min (OFF-band) of integration time 
are shown in Figs. \ref{color}, \ref{R} and \ref{net}. Fig. \ref{color} 
is a synthesized-color image whose Red, Green and Blue channels were given to the $r'$, 
ON-band and NET $\rm H\alpha$ frames respectively. 
Fig. \ref{R} shows the red continuum frame. Fig. \ref{net} shows the NET (ON - OFF-band) frame. 
The sky noise in all frames increases noticeably western of $\rm RA=11^h44^m48^s$, 
the edge of the WFC image, where the exposures consist only of the 2 hours taken at SPM.

\subsection{Spectroscopy}

The redshift of 5 dwarf systems in the region were obtained in February 2002 and in January 2003 using
the BFOSC spectrograph attached to the Cassini 1.5m telescope of the University of Bologna located at Loiano.
Spectra of another 4 dwarf systems were obtained 
in April 2002 using the CAFOS spectrograph attached to the 2.2m telescope at Calar Alto (Spain). 
Both spectrographs were equipped with low dispersion ($R \sim 500-800$) grisms.
In spite of the small telescopes used, the strong H$\alpha$ line prompted us with the redshift of these 
faint objects in short (15-20 min) exposures. 
Objects Dw1, Dw2, K5 and 97-114b show [OIII] ($\lambda 4959, 5007$ \AA) lines
in addition to H$\alpha$, but, unlike those obtained by Sakai et al. (2002) with the refurbished MMT, our 
1.5m spectra are of insufficient quality for measuring abundances.\\ 
Higher resolution spectroscopy was also acquired with BFOSC using a $R \sim 2300 $ grism
for two giant galaxies: CGCG97-125, CGCG97-114 and for the dwarfs Dw1 and Dw2 (see Fig. \ref{rotation}). 
The slit was positioned in the E-W direction except for Dw2 which was observed along the major axis
(P.A. 127 degrees).
The velocity of $\rm \sim 380 ~km~s^{-1}$, measured along the major axis of CGCG97-125,
likely corresponds to its total (corrected) rotational velocity.
Conversely the $\rm \sim 90 ~km~s^{-1}$ found for CGCG97-114 does not represent the
maximum rotational velocity since it was not taken along the galaxy major axis. 
The 2 dwarfs have observed total velocity widths of $\rm \sim 45$ and $\rm \sim 100 ~km~s^{-1}$ respectively, 
along the major axis. The edge-on corrections are however very uncertain for these objects.\\
The resulting rotation curves are shown in Fig. \ref{rotation} (filled dots for the giants, 
triangles for the dwarfs). 
The optical velocity curve of CGCG97-125 is in excellent agreement with the HI position-velocity
map of Sakai et al. (2002) (see their Fig. 8). 
Fig. \ref{rotation} also reports all other known redshifts (crosses) in the observed region 
(in the range $8000<Vel<8600$ $\rm ~km~s^{-1}$). 
The spatial axis corresponds to the projected distance (in arcsec) from the center of 97-125 along R.A.
Notice the monotonic velocity decrease with increasing projected separation from CGCG97-114 (see Section 4). \\
The results of the observations are summarized in Table 1 as follow:\\
Column 1: objects designations as given by Sakai et al. (2002).\\ 
Column 2, 3: celestial coordinates (with typical uncertainties of 2 arcsec).\\
Column 4: $r'$ band magnitudes.\\ 
Column 5: recessional velocities (with uncertainties).
The 4 redshifts in common with Sakai et al. (2002) are consistent with their measurements. 
Five redshifts, namely of K1, 97-114a and Dw2, K5 and 97-125a are given for the first time here.\\ 
Column 6, 7: $H\alpha+[NII]$ flux and E.W. obtained in this work.
Our $\rm H\alpha$ line photometry of five galaxies (CGCG97-125, CGCG97-114, Dw1, Dw2 and Dw3) 
in common with Sakai et al. (2002) is in full agreement with their measurements.\\ 
Column 8: star formation rate (SFR) derived from $\rm H\alpha$ flux (corrected for [NII] contamination and
for extinction) as in Kennicutt (1998).\\
Column 9: galaxy mass derived from $r'$ luminosity, transformed to  H band, 
applying the $M/L_H$ conversion of Gavazzi et al. (1996).\\

\section{Results}

\begin{figure}[!h]
\epsscale{1.0}
\plotone{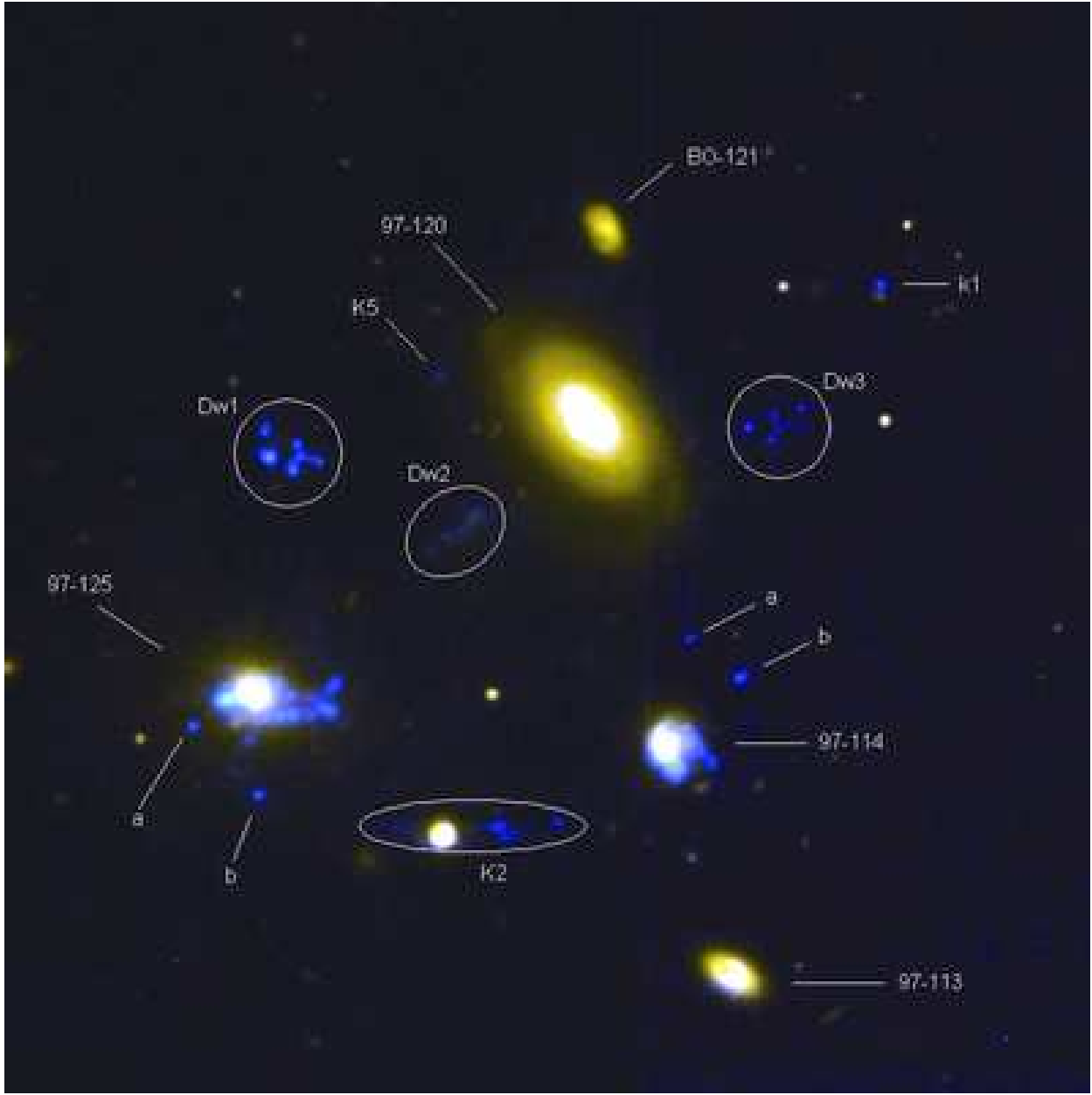}
\small{\caption{
The 5 $\times$ 5 arcmin$^{2}$ field centered on the group of galaxies projected near the X-ray center 
of Abell 1367. 
Blue corresponds to the $\rm H\alpha$ emission of hydrogen ionized by massive ($M>$10 M$\odot$) 
currently ($t<$20 Myr) forming stars. 
Along with two bright galaxies (CGCG97-114, CGCG97-125), 10 dwarf systems are simultaneously 
experiencing a burst of star formation. The yellow color corresponds to the red ($r'$ band) emission by old stars. 
All the $\rm H\alpha$ emitting objects are redshifted by ~1800$\pm$150 km~s$^{-1}$ with respect 
to the cluster systemic velocity, thus are belonging to a 
distinct group falling into the cluster from the front side.
}\label{color}}
\end{figure}

\begin{figure}[!h]
\epsscale{1.0}
\plotone{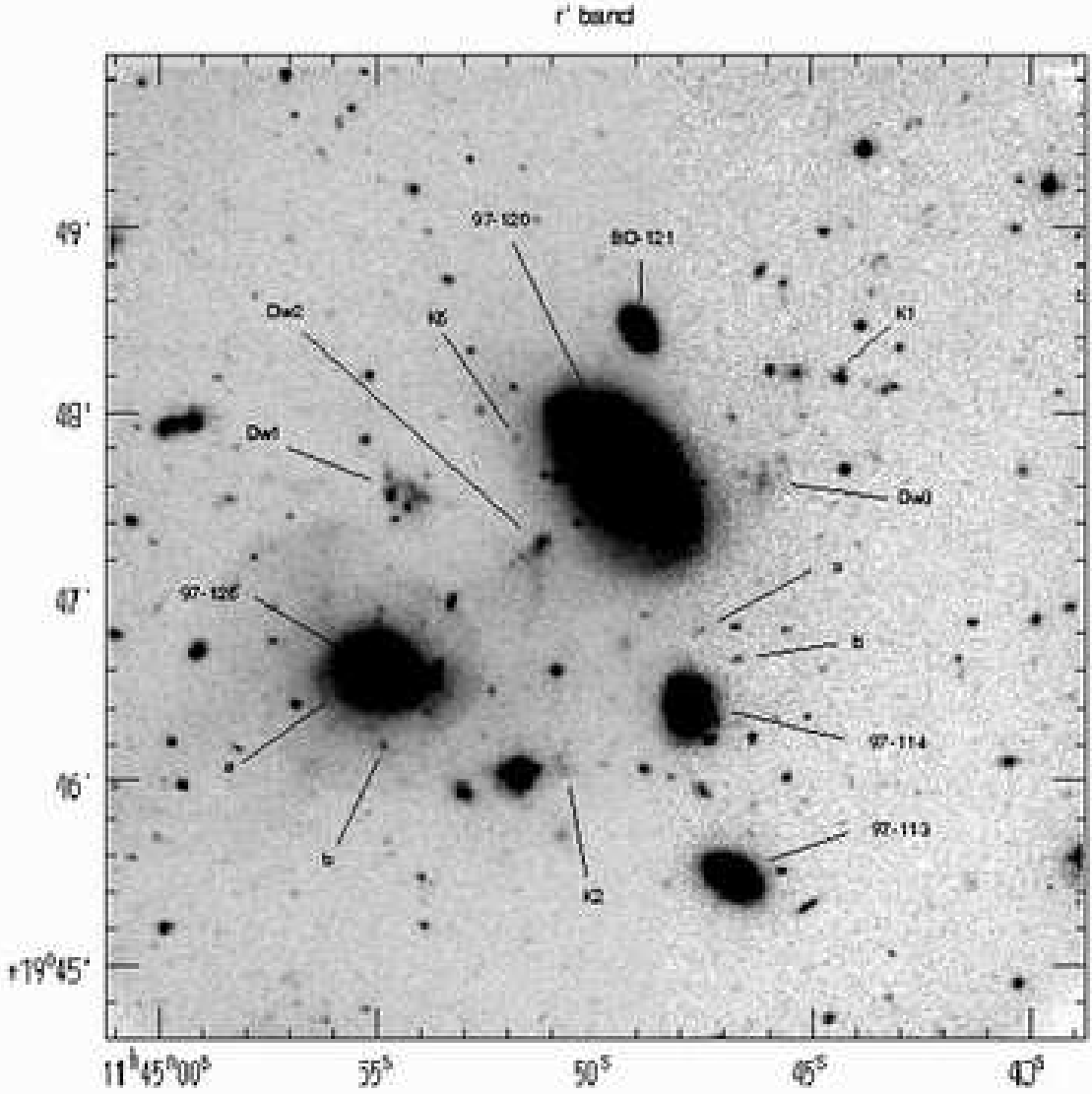}
\small{\caption{
High-contrast $r'$ band frame of the BIG group (with the J2000 celestial 
coordinate grid). All dwarf galaxies detected in the $\rm H\alpha$ frame have a visible 
counterpart in this red-frame, indicating that they contain old ($t >$ 3 Gyr) stars, 
in addition to the currently forming ones. 
The galaxy CGCG97-125 is surrounded by stellar-shells witnessing a merging event.
}\label{R}}
\end{figure}

\begin{figure}[!h]
\epsscale{1.0}
\plotone{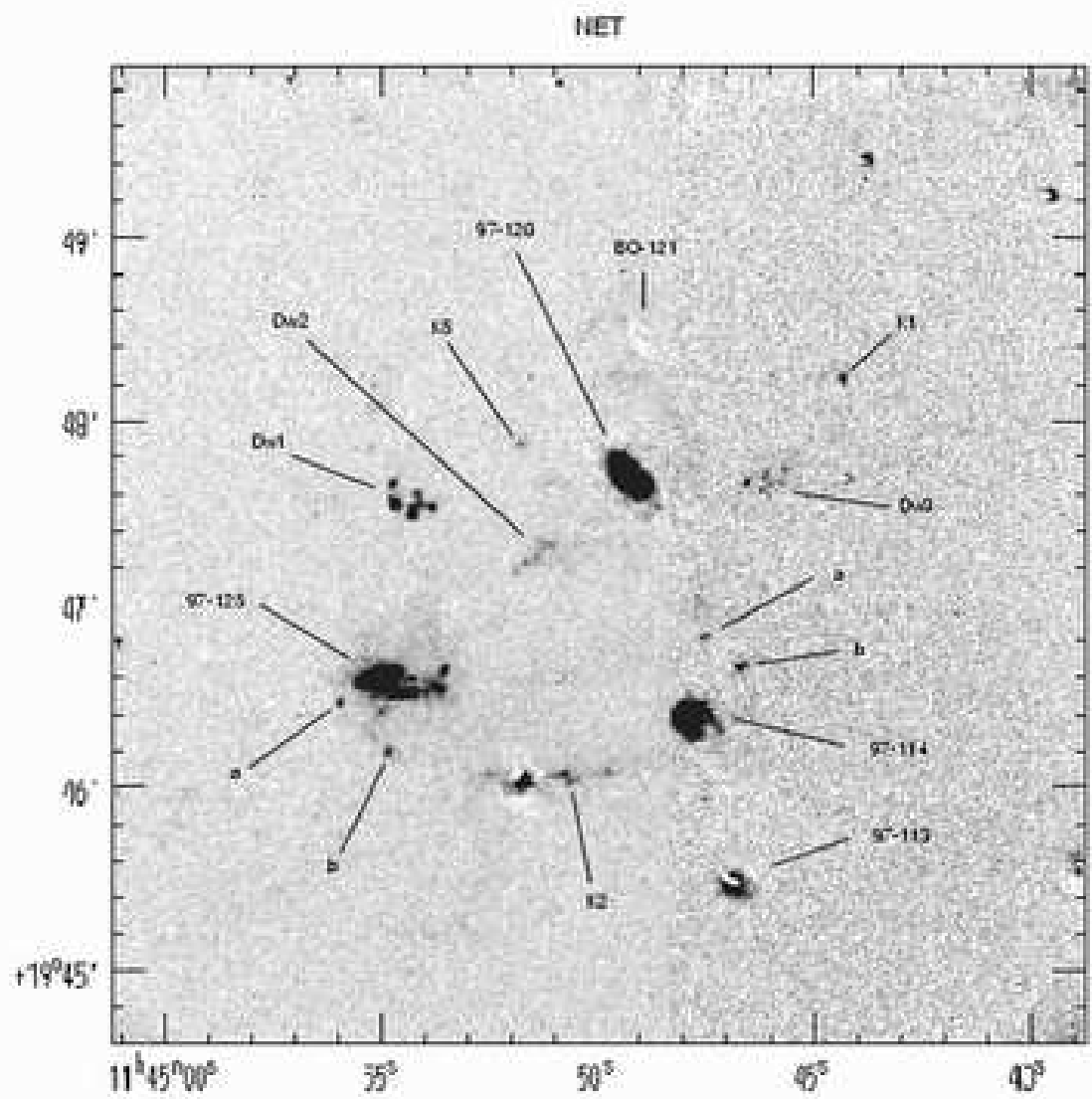}
\small{\caption{ Net ($\rm H\alpha + N[II]$) band frame of the BIG group.}\label{net}}
\end{figure}

\begin{figure}[!t]
\epsscale{1.0}
\plotone{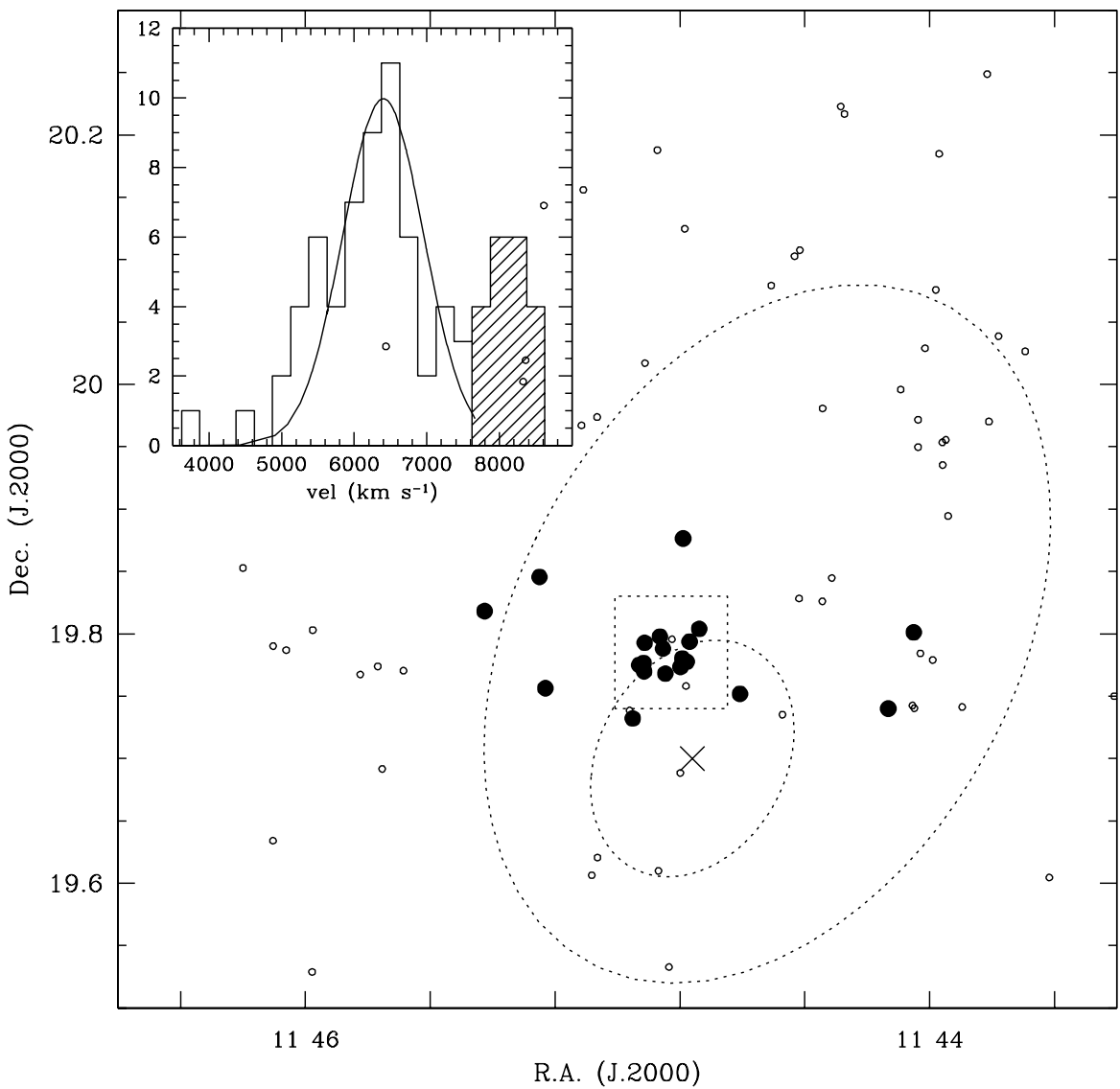}
\small{\caption{
Celestial distribution of galaxies with known recessional velocities in the 
cluster of galaxies A1367, with two contours of the X-ray emission sketched around the 
X-ray center (X). Small symbols mark galaxies within the gaussian velocity distribution of 
the cluster (see inset). Larger dots refer to galaxies with redshift exceeding 7700 km~s$^{-1}$, i.e. 
in the high velocity tail of the distribution (shaded). Notice that these galaxies are 
clustered in a small region, revealing the existence of a group falling onto the cluster. 
Galaxies inside the small square box (BIG), corresponding to Fig.\ref{color}, 
have been all detected in $\rm H\alpha$.
}\label{vel}}
\end{figure}

The high sensitivity of the present H$\alpha$ observations (all galaxies with H$\alpha$ luminosity in 
excess of $10^{38.5}$ erg~s$^{-1}$ were detected), allowed us to confirm the exceptional over-density of 
star forming galaxies projected near the X-ray center of the cluster of galaxies Abell 1367 
(see Fig.\ref{color}) independently reported by Sakai et al. (2001, 2002) and by Iglesias-Paramo et al. (2002).
Ten out of twelve star forming regions are associated with dwarf systems (or extragalactic HII regions) 
with H$\alpha$ equivalent widths often exceeding 100 \AA~(see Tab. 1). 
These galaxies, in spite of being $\sim 1000$ times smaller than typical giant galaxies, 
are currently forming stars at 10 times higher rate (per unit mass) than normal galaxies of similar luminosity, 
as derived from their $L(H\alpha)$ (Kennicutt 1998). As remarked by Sakai et al. (2002), it is the first time that 
such high density of star forming galaxies has been seen in a nearby cluster, 
in spite of having collected data over an area of A1367, Coma and the Virgo cluster approximately 500 times larger 
than the one shown in Fig.\ref{color}.\\
The H$\alpha$ dwarfs are found at the redshift of 8200$\pm$150 $\rm km~s^{-1}$ (see Table 1), 
consistent with the redshift of two bright galaxies (97-114 and 97-125) 
that was known from previous optical/radio spectroscopy. 
This velocity exceeds significantly the mean cluster velocity of $<V>=6420$ ($\sigma_V=822$)  $\rm km~s^{-1}$ 
(Struble \& Rood 1991). 
In addition to 12 galaxies in Fig.\ref{color}, redshifts in excess of 7700 $\rm km~s^{-1}$ are found for 
another 8 galaxies in A1367. They are all spatially segregated near the cluster center, 
and form a secondary peak in the cluster velocity distribution of Fig. \ref{vel}.
We consider unlikely that the group sits in the background of Abell 1367, in free Hubble flow
for two reasons. Firstly, at 30 arcmin projected angular separation
from the cluster center (see Fig. \ref{vel}) the group velocity of 8200 $\rm km~s^{-1}$ falls within
the caustics associated with the density enhancement of A1367 (see Fig. 7 of Gavazzi, Randone \& Branchini 1995). 
In other words the group is well within the turn-around radius of the cluster.
This points out the existence of a separate dynamical unit falling into the 
cluster from the near side. The compact part of this group, comprising 12 central group members, 
will be referred to as the "blue infalling group" (BIG) hereafter.
Three objects are not associated with BIG (see Table 1): CGCG97-113 is the only galaxy in the region
with a velocity within the velocity dispersion of A1367; BO[121] lies in the background; 
the third massive galaxy NGC 3860 (marked 97-120 in Fig. \ref{color}) ($\rm V_{hel}= 5635$ $\rm km~s^{-1}$) 
is blueshifted with respect to A1367 by approximately 800 $\rm km~s^{-1}$.
Observations of the neutral hydrogen (HI) line (Chincarini et al. 1983) show that 
NGC 3860 has lost approximately 90 \% of its original hydrogen content ($\rm HI_{def}$=0.9),  
indicating that the galaxy has crossed the cluster core, and that the ram-pressure (Gunn \& Gott 1972) 
exerted by the dense intergalactic medium (IGM) (with a central 
density $\rho_{IGM}$ = 1.25~10$^{-3}$ atoms~cm$^{-3}$) 
(Mohr et al. 1999) has caused its hydrogen deficiency. \\
Conversely 97-125, the brightest member of BIG, has a normal 
hydrogen content ($\rm M_{HI} = 3.9 \times 10^9 M_\odot$ (Sakai et al. 2002), implying  $\rm HI_{def}$=-0.21), 
pointing out that this galaxy has not yet crossed the dense cluster medium. 
However its HI column density distribution appears asymmetric, with the highest signal in the western 
side of the galaxy, as seen in
HI map obtained by Sakai et al. (2002) (see their Fig. 7). 
Although the asymmetry is interpreted by Sakai et al. (2002) as due to a tidal interaction 
among BIG members, it is also consistent with the hypothesis that the galaxy is entering the 
cluster IGM and its HI content is being removed from the eastern side and accumulated in the western side, 
possibly due to ram-pressure, as found in other infalling galaxies discussed by Gavazzi (1989).
Furthermore the HI map of Sakai et al. (2002) reveals a marginal detection associated with GCGC97-114, 
with a resulting HI deficiency of 0.7. This high value suggests that
this galaxy is currently undergoing gas erosion by ram-pressure, thus it has already encountered the cluster IGM.
We conclude that the BIG galaxies are plausibly at the front-edge of A1367, 
beginning their journey through the cluster IGM.\\
Unfortunately the Tully-Fisher (1977) distance estimator is not accurate enough to confirm this 3-D picture.
With this method we obtain a distance of 88 $\pm 25$ 
and 82 $\pm 25$ Mpc for GCGC97-125 and GCGC97-114 respectively, consistent
with the distance of 85 Mpc of A1367, but not sufficiently accurate to exclude that
BIG is in Hubble flow at 110 Mpc.\\ 
All star-forming galaxies in BIG have detectable continuum emission in the red-band image 
shown in Fig. \ref{R}. This demonstrates that they all contain a significant 
($10^{7}-10^{9}$ M$\odot$, as estimated from their $r'$ luminosity using the M/L conversion of 
Gavazzi et al. 1996) population of old ($t > 3$ Gyr) stars, in addition to their young ($t < 20 $ Myr) 
stars revealed in H$\alpha$. 
The $r'$ band luminosity function of BIG in the interval -19.2 $< M_{r'} <$ -12.2 has a 
slope of only $\alpha_{r'}$ = -1, 
shallower than $\alpha_r$  = -1.20 found in the deepest cluster surveys (Trentham \& Tully 2002). 
It is consistent with 
the $r'$ band luminosity function of Hickson compact groups (Hunsberger et al. 1998), 
suggesting that originally BIG was a normal compact group. 
On the opposite, BIG represents a true exception as far as its young stars content. How this translates in the 
slope of the H$\alpha$ luminosity function? The existing H$\alpha$ observations of compact groups 
(Iglesias-Paramo \& Vilchez 1999; Vilchez \& Iglesias-Paramo 1998) are unfortunately 
10 times shallower than the present ones, 
making their $\alpha_{H \alpha}$ undetermined at the faint levels of the present survey. 
Thus we derived the H$\alpha$ luminosity function of BIG and compared it with the one of the cluster Abell 1367 
(Iglesias-Paramo et al. 2002). 
The faint end of BIG has a slope $\alpha_{H \alpha}$ = -1.25, significantly steeper than $\alpha_{H \alpha}$ = -0.8 
found in the cluster. In other words the ratio of faint-to-bright galaxies is 5 times 
larger in this group than in the cluster. \\
 
\section{Discussion}
We have shown that the H$\alpha$ glowing galaxies projected near the center of A1367
are relatively "old" galaxies that would have 
escaped attention when observed in broad-band images. Similar overdensities of star forming objects 
have never been observed neither in nearby clusters nor in compact groups. 
What is the mechanism that triggers the burst of star formation, for the first time seen here, 
in an otherwise normal group falling into a rich cluster of galaxies? Is the perturbation 
locally excited, or induced by an external agent?\\

\subsection{Tidal Interaction among group galaxies}

Sakai et al. (2002) favour the internal interpretation. They argue that the burst of star formation is caused by 
tidal encounters between the group members. They proposed that the dwarf systems in BIG formed from
enriched material stripped from one of the larger galaxies due to a recent tidal interaction.
Their argument is supported by three evidences:
i)  The stellar-shells surrounding the 
galaxy 97-125 at approximately 30, 45 and 60 arcsec from its center (see Fig. \ref{R}), indicate that it has 
merged with another group galaxy, as shells around galaxies result either from minor (Thomson \& Wright 1990) or major
(Schweizer 1980)   
merging events. The H$\alpha$ filament observed within the stellar envelope of 97-125 might be the 
remnant of a recently swallowed galaxy.
ii) The metallicity of one of the faint knots (K2) is 
similar to the one of 97-125, i.e. approximately solar. This rules out the evolutionary scenario
in which the intergalactic knots K2 are isolated normal dwarf galaxies evolving as closed systems,
because such faint objects are expected to have metallicities in the range 1/20 - 1/40 solar.
iii) The presence of an HI bridge extending through many of the emission-line regions, 
argues in favour of the tidal interpretation.\\

There are arguments against this interpretation: 
i) Although there is clear evidence of merging in CGCG97-125 (shells), we argue that the
overall morphology of BIG does not fully support the tidal interaction picture  
because no evidence of large-scale tidal tails is found here, providing the clear signature
of a merging event (see the Antennae in Hibbard et al. 2001 and the simulations of Barnes \& Hernquist 1992). 
ii) Should the abnormal star formation in BIG be triggered by tidal interactions among its galaxies, 
similar phenomena would be frequently observed in other compact groups, contrary to the
observations of Vilchez \& Iglesias-Paramo (1998).
Among BIG dwarfs, only Dw1, Dw2 and Dw3 have broad-band colors (B-R) available from Sakai et al. (2002)
(0.05, 0.53 and 0.2 respectively). These colors are marginally bluer than the average $<B-R>=0.75\pm0.4$
found among 143 tidal dwarf candidates in 10 interacting systems studied by Weilbacher et al. (2000).
This blue excess confirms the exceptional nature of BIG among other compact groups 
even those containing tidal interacting systems.
iii) While the faint knots (K2) show a high metallicity, similar to the one of 97-125, 
the metal abundances of the two other dwarf systems Dw1 and Dw3 (Sakai et al. 2002) 
are consistent with their being irregular galaxies evolved independently.\\ 

\subsection{Ram-pressure interaction with A1367}

\begin{figure}[!t]
\epsscale{1.0}
\plotone{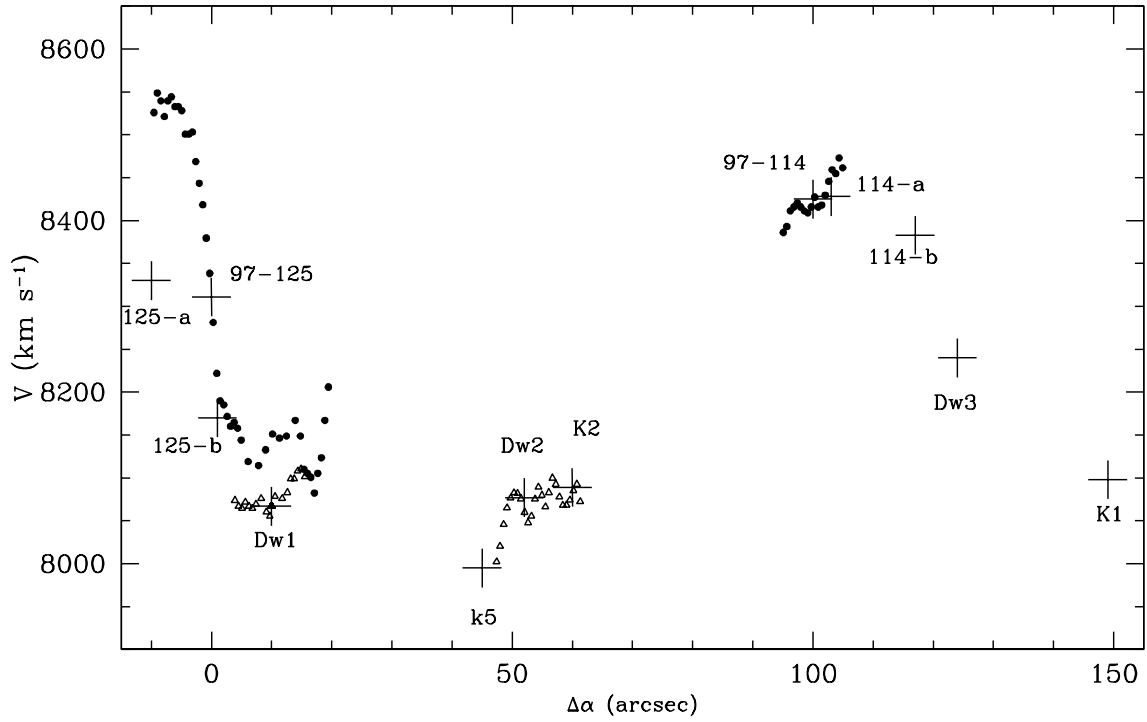}
\small{\caption{Distribution of the known recessional velocities (crosses) as a function of the projected 
distance (in arcsec) from the center of 97-125 along R.A.
Rotation curves of 97-125 and 97-114 are given with filled dots, those of Dw1 and Dw2 with triangles.}\label{rotation}}
\end{figure}

We offer an alternative interpretation emphasizing that BIG is likely 
infalling onto A1367 at high speed. \\
Galaxy infall occours elsewhere onto this cluster. Gavazzi et al. (2002) found evidence for enhanced current star
formation activity associated with faint ($\rm -17<M_p<-19 ~mag$) galaxies, consistent with these
being objects recently "captured" by the cluster. Similarly Bravo-Alfaro et al. (2000, 2001)
found that several spiral galaxies in the Coma cluster have their neutral hydrogen content significantly displaced
from their optical centroids, consistent with their recent arrival into the cluster hostile environment.  
Among galaxies in A1367, CGCG97-073 and CGCG97-079 were studied in some detail by Gavazzi et al. (1995; 2001)
who found that both galaxies have their present star formation enhanced along peripheral HII regions which developped 
at the side facing the direction of motion through the dense cluster IGM. 
There are indications that their neutral hydrogen is displaced in the opposite side (Dickey \& Gavazzi 1991). 
Trails of 50 kpc length are detected  both in the light of the synchrotron radiation 
(Gavazzi et al. 1995) and in H$\alpha$ (Gavazzi et al. 2001) supporting the idea that ram-pressure (Gunn \& Gott 1972) 
is for a limited amount of time 
enhancing the star formation of galaxies that are for the first time entering the cluster medium.\\ 
During their travel at 1800 $\rm km~s^{-1}$ through the IGM, whose density gradually 
increases from $\rho_{IGM}$ $\sim$ 10$^{-4}$ atoms~cm$^{-3}$ at their present peripheral location, to 
$\rho_{IGM}$ = 1.25~10$^{-3}$ atoms~cm$^{-3}$ near the center of A1367, the external ram-pressure 
experienced by the galaxies of BIG will increase  from:
\begin{displaymath}
P = \rho v^2 \sim 5~10^{-12} [\rm dyn ~cm^{-2}] 
\end{displaymath}
to an order of magnitude higher.\\
The galaxies in BIG have low stellar surface densities 
($\sigma_S$ $\sim$ 5 10$^{-3}$ g~cm$^{-2}$) and interstellar gas surface densities 
($\sigma_g$ $\sim$ $10^{-3}$ g~cm$^{-2}$) typical of dwarfs galaxies. 
Thus the restoring gravitational force (pressure) at their interior: 
\begin{displaymath}
F = 2 \pi G \sigma_S  \sigma_g \sim 2~10^{-12} [\rm dyn ~cm^{-2}]
\end{displaymath}
results significantly smaller than the ram-pressure.
In conclusion the external pressure is sufficient to compress and shock their interstellar gas,
thereby triggering a star formation burst, as suggested by Fujita \& Nagashima (1999).
In the long run the increasing ram-pressure will fully strip their gaseous 
material leading to the complete ablation of their interstellar gas, thus suppressing the star formation due to 
fuel exhaustion. 
The stripped blobs of typical radius $R$ of 5 kpc and mass $M$ = $10^8 M_{\odot}$ might even experience a deceleration:
\begin{displaymath}
a = \frac {(P - F) \pi  R^2}{M} = 1.5 ~10^{-8} [\rm cm ~s^{-2}]
\end{displaymath}
with a consequent measurable velocity decrease of $\Delta V=300~ \rm km ~s^{-1}$ in a time as short as
$10^8$ yrs.
Such an acceleration could explain the monotonic velocity decrease 
observed in Fig. \ref{rotation} in blobs (97-114a, 97-114b, Dw3, K1) of increasing projected distance
from CGCG97-114.\\ 
The weaknesses of this second hypothesis, as pointed out by Sakai et al. (2002) are: 
i) none of the galaxies in BIG show evidence of the bow-shock structure in the H$\alpha$ or broad-band
images that is characteristic of the other objects of this type in A1367.
ii) Chandra observations (Sun \& Murray 2002) do not show soft X-ray extended emission, contrary to the
expectations. However we argue that the region in
Fig.1 was not fully covered by Chandra, lying near the edge of chip "S3". CGCG97-125 and Dw1 were out of the 
observed field while CGCG97-114 and Dw3 were in fact detected as extended and point-like sources respectively.\\
In conclusion, we agree with Sakai et al. (2002) that tidal interactions among members of the BIG compact group 
must have had a role in triggering the burst of star formation seen in this group for the first time.
The stellar shells around its brightest galaxy (CGCG97-125) clearly testify such an event and the quasi solar 
metallicity of knots K2
argues in favour of this object being composed of evolved material stripped from CGCG97-125.\\
We remark that the high velocity infall of the group onto A1367   
might have enhanced the star-bursting transient phase making BIG unique among other isolated
compact groups of galaxies.

\acknowledgements

We wish to thank G. Chincarini, M. Colpi, L. Mayer and B. Poggianti for stimulating discussions. 
L. Carrasco research is supported by CONACYT research grant G28586-E. This research has made use of the 
NASA/IPAC Extragalactic Database (NED) and of the GOLDmine database. NED is operated 
by the Jet Propulsion Laboratory, California Institute of Technology, under contract with the
National Aeronautics and Space Administration. 
GOLDmine is operated by the Universita' degli Studi di Milano-Bicocca.

\begin{landscape}
\begin{table}
\label{Tab1}
\caption {Parameters of the galaxies in the BIG group. Objects are labelled following Sakai et al. (2002)}
\[
\begin{array}{p{0.1\linewidth}rcccccccc}
\hline
\noalign{\smallskip}  
Name & R.A. & Dec & r' &  Vel  &  F(H\alpha+[NII]) & EW(H\alpha+[NII]) & SFR & LogMass \\
     & (J2000) & (J2000) &  mag   & km~s^{-1} &   erg~cm^{-2}~s^{-1} & \AA 
     & M_{\odot}~yr^{-1}  &  M_{\odot}\\
 (1) & (2) & (3) & (4) &  (5) & (6) & (7) & (8) & (9)\\ 
\noalign{\smallskip}
\hline
\noalign{\smallskip}
K1            &   11 44 44.28 & 19 48 14.00 &  20.9  & 8098 \pm 80\tablenotemark{a}~  & 6.92 (\pm 1.73) 10^{-16} & 79\pm 10   & 0.005 & 7.8  \\
Dw3           &   11 44 46.13 & 19 47 37.50 &  19.3  & 8240 \pm 77\tablenotemark{a,b}~ & 3.19 (\pm 0.90) 10^{-15} & 88\pm 24   & 0.023 & 8.6  \\
97-113        &   11 44 46.60 & 19 45 29.00 &  15.2  & 6419 \tablenotemark{c}~ &    -    & -    & -     & 10.8 \\
97-114b       &   11 44 46.68 & 19 46 39.50 &  22.0  & 8383 \pm 62\tablenotemark{a,b}~  & 1.28 (\pm 0.20) 10^{-15} & 546\pm 26   & 0.009 & 7.2  \\
97-114a       &   11 44 47.54 & 19 46 48.80 &  21.8  & 8428 \pm 60\tablenotemark{a}~ & 7.70(\pm 1.70) 10^{-16}  & 222\pm 17   & 0.006 & 7.3  \\
97-114        &   11 44 47.88 & 19 46 24.60 &  15.4  & 8425 \pm 32\tablenotemark{a,c}~ & 5.28(\pm 0.88) 10^{-14}  & 38\pm 3     & 0.380 & 10.6 \\
BO121         &   11 44 48.90 & 19 48 29.00 &  16.2  & 20296 \tablenotemark{c}~ &     -   & -    & -     & 11.2 \\
97-120        &   11 44 49.51 & 19 47 44.10 &  13.6  & 5635 \pm 33\tablenotemark{a,c}~ & -    & -            & - & 11.6 \\
K2            &   11 44 50.81 & 19 46 05.10 &  20.2  & 8089 \pm 70\tablenotemark{a,b}~ & 2.98(\pm 0.38) 10^{-15}  & 184\pm 8     & 0.021 & 8.2  \\
Dw2           &   11 44 51.27 & 19 47 17.50 &  19.2  & 8077 \pm 61\tablenotemark{a}~ & 2.13(\pm 0.54) 10^{-15} & 52\pm 15   & 0.015 & 8.7  \\
K5            &   11 44 51.84 & 19 47 51.70 &  22.7  & 7995 \pm 60\tablenotemark{a}~ & 6.52(\pm 1.71) 10^{-16} & 502\pm 89   & 0.005 & 6.8  \\
Dw1           &   11 44 54.22 & 19 47 33.20 &  18.3  & 8067 \pm 59\tablenotemark{a,b}~  & 1.00(\pm 0.11) 10^{-14} & 108\pm 6     & 0.072 & 9.1  \\
97-125b       &   11 44 54.89 & 19 46 11.30 &  21.7  & 8170 \tablenotemark{b}~     & 8.56(\pm 1.81) 10^{-16} & 218\pm 31   & 0.006 & 7.4  \\
97-125        &   11 44 54.99 & 19 46 35.80 &  14.5  & 8311 \pm 35\tablenotemark{a,c}~  & 8.47(\pm 1.10) 10^{-14} & 25\pm 2     & 0.610 & 11.1 \\
97-125a       &   11 44 55.99 & 19 46 28.00 &   -    & 8330 \pm 50\tablenotemark{a}~  & -  &  -     &  - &  - \\

\noalign{\smallskip}
\hline
\end{array}
\tablenotetext{a}{this work}
\tablenotetext{b}{Sakai et al. (2002)}
\tablenotetext{c}{NED}
\]

\end{table}
\end{landscape}

\smallskip
\onecolumn

\end{document}